\documentstyle[12pt,fleqn,epsf,epsfig]{article}

\begin{document}
\title{Robust and Irreversible Development in Cell Society as a
General Consequence of Intra-Inter Dynamics}
\author{Kunihiko Kaneko and Chikara Furusawa\\
{\small \sl Department of Pure and Applied Sciences,
College of Arts and Sciences,}\\
{\small \sl University of Tokyo,}\\
{\small \sl Komaba, Meguro-ku, Tokyo 153, Japan}\\
}

\date{}




\maketitle

\begin{abstract}

A dynamical systems scenario for developmental cell biology is proposed, based on 
numerical studies of a system with interacting units with internal dynamics and
reproduction.  Diversification, formation of discrete and recursive types, and
rules for differentiation are found as a natural consequence of such a system.
``Stem cells" that either proliferate or differentiate to different types stochastically 
are found to appear when intra-cellular dynamics are chaotic. 
Robustness of the developmental process against microscopic and macroscopic
perturbations is shown to be a natural consequence of such intra-inter dynamics, while
irreversibility in developmental process is discussed in terms of the gain of stability,
loss of diversity and chaotic instability.  

\end{abstract}

\section{Intra-inter Dynamics for Cell Biology}

A biological unit has always some internal structure that can change in time.
These units, however, are not separated from the outside world completely.
For example, consider a cell.
In this case, isolation by a biomembrane is flexible and incomplete.
When the units are represented by dynamical systems,
they interact with each other and with the external environment.
Hence, we need a model consisting of the interplay between inter-unit and intra-unit dynamics 
\cite{KK90,KK93,KK98a}.  Here we will mainly
discuss the developmental process of a cell society accompanied by
cell differentiation, where
the intra-inter dynamics consist of several biochemical reaction
processes.  The cells interact through the diffusion of
chemicals or their active signal transmission.  

In this case ( and generally in a biological system), the number of elements is not
fixed in time, but they are born (and die) in time.
Consider development of cell society with cell divisions.
After the division of a cell, if two cells remained identical, 
another set of variables would not be necessary.
If the orbits have orbital instability (such as chaos), however,
the orbits of the (two) daughters will diverge.
Thus the increase in the number of variables is tightly connected with the
internal dynamics.

We have studied several models\cite{KY94,KY97,KY98,KK97,FK98a,FK98b} choosing
(a) the internal variables and their dynamics,
(b) interaction type, and
(c) the rule to change the degrees of freedom (e.g., cell division).

As for the internal dynamics, auto-catalytic reaction among
chemicals is chosen. Such auto-catalytic
reactions are necessary to produce chemicals in a
cell, required for reproduction\cite{Eigen}.  
Auto-catalytic reactions often lead to nonlinear oscillation in chemicals.
Here we assume the possibility of such
oscillation in the intra-cellular dynamics
\cite{Goodwin,Hess}.
As the interaction mechanism, the diffusion
of chemicals between a cell and its surroundings is chosen.

To be specific, we mainly consider the following model here.
First, the state of a cell $i$  is assumed to be characterized by the cell volume and
a set of functions $x^{(m)} _i(t)$ representing the concentrations of $k$ chemicals
denoted by $m=1,\cdots,k$.  The concentrations of chemicals change as a result of
internal biochemical reaction dynamics within each cell and
cell-cell interactions communicated through the surrounding medium.

For the internal chemical reaction dynamics, we choose a catalytic network among
the $k$ chemicals.  The network is defined by a collection of triplets ($\ell$,$j$,$m$)
representing the reaction from chemical $m$ to $\ell$  catalized by $j$.
The rate of increase of $x^{\ell}_i(t)$ (and decrease of $x^{m}_i(t)$)
through this reaction is given by $x^{(m)}_i(t) (x^{(j)}_i(t))^{\alpha}$, where
$\alpha$ is the degree of catalyzation
($\alpha = 2$ in the simulations considered presently).
Each chemical has several paths to other chemicals, and thus a complex
reaction network is formed.  The change in the chemical concentrations through all such
reactions, thus, is determined by the set of all terms of
the above type for a given network.
These reactions can include genetic processes, where $x^{\ell}_i(t)$ is
regarded as expression of a given gene.

Cells interact with each other through the transport of chemicals out of and into
the surrounding medium.  As a minimal case, we consider only indirect
cell-cell interactions through diffusion of chemicals via the medium.
The transport rate of chemicals into a cell is proportional to
the difference in chemical concentrations between the inside and the outside of
the cell, and is given by $D(X^{(\ell)} (t) -x^{({\ell})}_i (t))$,
where $D$ denotes the diffusion constant, and $X^{(\ell)} (t)$ is the concentration of
the chemical at the medium. 
The diffusion of a chemical species through cell membrane should depend on
the properties of this species.
In this model, we consider the simple case in which there are two types of
chemicals, one that can penetrate the membrane and one that cannot.
With this type of interaction, corresponding chemicals in the medium are consumed.
To maintain the growth of the organism, the system is considered to be immersed
in a bath of chemicals through which (nutritive) chemicals are supplied to the cells.

As chemicals flow out of and into the environment,
the cell volume changes.  The volume  is assumed to be proportional 
to the sum of the quantities of chemicals in the cell, and thus is a dynamical variable.
Accordingly, chemicals are diluted as a result of the increase of the cell volume.

In general, a cell divides according to its internal state, for example,
as some products, such as DNA or the membrane, are
formed, accompanied by an increase in cell volume.  Again,
considering only a simple situation, we assume that a cell divides into two
when the cell volume becomes double the original.
At each division,  all chemicals are almost equally divided,
with random fluctuations.

Of course, there can be a variety of choices on the
chemical reaction network.  The observed results
do not depend on the details of the choice, as long as the
network allows for the oscillatory intra-cellular dynamics leading to the
growth in the number of cells.
Note that the network is not constructed to imitate
an existing biochemical network.
Rather, we try to demonstrate that important features in a biological system
are the natural consequence of a system with internal dynamics,
interaction, and reproduction.  From the study we extract a
universal logic underlying this class of models.

In this sense, our approach is constructive\cite{KT94,KK98b}.
We first construct a simple `world' by combining fundamental procedures, and
clarify universal class of phenomena.  Then we try to
reveal the underlying universal logic
that the life process should obey.
With this we try to provide a new look at the present organisms
from our discovered logic. 

We believe that this `constructive' approach is essential to understand the
logic of life.  As long as we study only the present organisms in nature,
it is hard to distinguish the logic that organisms
necessarily should obey, from frozen accident through evolution.
The logic and universality class
can be clarified only by constructing some world.
The present organisms, then, are understood as
one representative of a universal class, to which the
life-as-it-could-be also belongs.

{\sl Remarks}

In the pioneering study, Turing proposed a
pattern formation mechanism through spatial symmetry breaking 
induced by dynamic instability \cite{Turing,Newman}. 
On the other hand, existence of multiple attractors in genetic networks are found
that could be corresponded to different cell types\cite{Kauffman}.
In our approach the generation of differentiation
rules is studied by introducing a rich internal dynamics
(including chaos), and the reproduction of units.   With this reproduction,
initial conditions are selected to form discrete cell types and
rules for differentiation.

\section{Scenario for Cell Differentiation}

From several simulations of the models
starting from a single cell initial condition, Yomo and the authors
proposed the following scenario (``isologous diversification"), as
a general mechanism of spontaneous differentiation of replicating
biological units \cite{KY97,KY98}.  First, we briefly review this scenario.
(see Fig.1 for schematic representation):

1)Synchronous divisions with synchronous oscillations of the chemicals

Up to a certain number of cells (which depends on the model parameters),
dividing cells from a single cell have the same
characteristics.  Although each cell division is not exactly identical due
to the accompanying fluctuation in the biochemical
composition, the phase of oscillation in the concentrations, and as a result,
the timing of cell division, remains synchronous for all cells.
The synchronous cell division is also observed with the cells in the 
embryo-genesis of mammals  up to the eight cells.  

2)Clustering in the phases of oscillations

  When the number of cells rises above a certain (threshold) value,
the state with identical cells is no longer stable.
Small differences introduced by the fluctuation
start to be amplified, until the synchrony of the 
oscillations is broken.  Then the cells split into a few groups,
each having a different
oscillation phase.  The cells belonging to each group are
identical in phase.  This diversification in the phases, however, 
cannot be called
cell differentiation, because  the time average of the biochemical 
concentrations reveals that the cells are almost identical.
The change of phases at the second stage is due to dynamic clustering,
studied in coupled nonlinear oscillators \cite{KK90,KK94b}. 
Clusters are formed through the balance between the two tendencies, and
the number ratio of the clusters satisfies some condition, to have
stability against perturbations.

3)Differentiation in chemical composition

        With the further increase of the cell number,
the average concentrations of the biochemicals over the cell
cycle become different.  The composition of
biochemicals as well as the rates of catalytic reactions and transport of
the biochemicals become different for each group.
The orbits of chemical dynamics plotted in the
phase space of biochemical concentrations,
lie in a distinct region within the phase space, while
the phases of oscillations remain different by cells within each group
(see Fig. 1b).

Hence  distinct groups of cells are formed with different chemical characters.
Each group is regarded as a different cell type,
and the process to form such types is called differentiation.
With the nonlinear nature of the reaction network, the difference in chemical
composition between the clusters is amplified.  By the formation
of  groups of different chemical compositions, each
intra-cellular biochemical dynamics is again stabilized. 
Some internal degrees of freedom are necessary to support the difference in
the phase space position at this stage.  

4)Determination of the differentiated cells

After the formation of cell types, the chemical compositions
of each group are inherited by their daughter cells. In other words,
chemical compositions of cells are recursive over divisions.
The biochemical properties of a cell are inherited by its progeny, or in other
words, the properties of the differentiated cells are stable, fixed or
determined over the generations (see Fig. 1c).  
After several divisions, such initial condition of units is
chosen  to give the next generation of the same  type as its mother cell.

5)Generation of rules for hierarchical organization

As the cell number increases, further differentiation proceeds.
Each group of cells differentiates further into two (or more) subgroups (See Fig.2).
Thus, the total system consists of
units of diverse behaviors, forming a heterogeneous society.

The most interesting example here is the formation of
stem cells, given in Fig.2 \cite{FK98a}. This cell type, denoted as `S' here, 
either reproduces the same type or forms different cell types,
denoted for example as type A and type B.  Then after division
events $S \rightarrow S,A,B$ occur.  Depending on the adopted chemical
networks, the types A and B replicate, or switch to different 
types.  For example $A \rightarrow A, A1,A2,A3$ is observed in
some network.  This hierarchical organization is
often observed when the internal dynamics have some complexity,
such as chaos.

The differentiation here is ``stochastic", arising from chaotic 
intra-cellular chemical dynamics.
The choice for a stem cell either to replicate or to differentiate
looks like stochastic as far as the cell type is
concerned.  Since such stochasticity
is not due to external fluctuation but is a result of the internal state,
the probability of differentiation can be regulated by the intra-cellular state.
This stochastic branching is accompanied by
a regulative mechanism.  When some cells are removed externally during 
the developmental process, the rate of differentiation changes so that
the final cell distribution is recovered.

In some biological systems such
as the hematopoietic system, stem cells  either replicate or
differentiate into different cell type(s).  This differentiation
rule is often hierarchical \cite{Alberts,Ogawa}.
The probability of differentiation to one of the several blood cell
types is expected to depend on the interaction.
Otherwise, it is hard to explain why the developmental
process is robust.  For example, when the number of some terminal cells
decreases, there should be some mechanism to increase the rate of
differentiation from the stem cell to the terminal cells.
This suggests the existence of interaction-dependent regulation of the
type we demonstrated in our results.

{\sl Remarks}

Of course the behavior of our cell dynamics depends on the specific choice of
biochemical networks\footnotemark .  Indeed the network with stem cells (with chaotic dynamics)
is not so common.  However, such cell system with stem cells and
differentiation generally has a higher growth speed as an ensemble of cells
\cite{FK99}. In this sense, it is natural to expect that a multicellular system with
such biochemical network that allows for chaotic stem cells has been selected
through evolution.

\footnotetext{ In some examples with a larger diffusion coupling, cells that may be termed 
`tumor-type cell' appear.  These cells differentiate in an extraordinary way,
destroy the cooperativity attained in 
the cell society, and grow faster in numbers in a selfish way.  
Their chemical configuration loses
diversity, and the ongoing chemical pathway there is simpler than other cell
types \cite{KY97}.}

\section{Stability}

In a developmental system, there are four kinds of stability,
which are mutually interrelated.

\begin{itemize}

\item
Microscopic:
The developmental process is
robust against molecular fluctuations.   

\item
Macroscopic:
The process is robust 
against macroscopic perturbations such as the removal of some cells or
somatic mutations.

\item
Path:
Not only the final states but also the
developmental path resulting in such states is stable against microscopic and 
macroscopic perturbations.

\item
Generation:
The developmental process is recursive to the next generation.  In a 
multicellular organism, the dynamical process from a single germ cell to 
adult organism is almost preserved to the
next generation, in spite of a long-term process amidst all the fluctuations involved therein.  

\end{itemize}

{\sl Microscopic Stability}

The developmental process is stable against molecular fluctuations.
First, intra-cellular dynamics of each cell type are stable against such
perturbations.  Then, one might think that this selection of each cell type is 
nothing more than a choice among basins of attraction
for a multiple attractor system.  If the interaction were neglected,
a different type of dynamics would be interpreted as a different attractor.
In our case, this is not true, and cell-cell interactions are necessary to 
stabilize cell types\footnotemark.  Given cell-to-cell interactions,
the cell state is stable against perturbations on the level of
each intra-cellular dynamics. 

\footnotetext{Importance of cell-cell interaction is discussed in the
study of equivalent cells \cite{equiv}  and community effect \cite{comm}.}

Next, the number
distribution of cell types is stable against fluctuations.  Indeed,
we have carried out simulations of our model, by adding a noise term,
considering finiteness in the number of molecules.
The obtained cell type as well as the number distribution is hardly affected
by the noise as long as the noise amplitude is not too large \cite{KY98}.
When the noise amplitude is too large, distinct types are no longer formed.
Cell types are continuously distributed.  In this case, the division speed 
is highly reduced,  since the differentiation of roles by differentiated 
cell types is destroyed.

{\sl Macroscopic Stability}

Each cellular state is also stable against perturbations of the interaction
term.  If the cell type number distribution is changed within some range, 
each cellular dynamics keep its type.  Hence discrete, stable
types are formed through the interplay between intra-cellular dynamics
and interaction. 
The recursive production is attained through the selection of initial
conditions of the intra-cellular dynamics of each cell, so that it is rather
robust against the change of interaction terms as well.

The macroscopic stability is clearly shown in the 
spontaneous regulation of differentiation ratio in the
last section.  How is this interaction-dependent rule formed?
Note that depending on the distribution of the other cell types,
the orbit of internal cell state is slightly deformed.
For a stem cell case,
the rate of the differentiation or the replication
( e.g., the rate to select an arrow among $S\rightarrow S,A,B$)
depends on the cell-type distribution.  For example,
when the number of ``A" type cells is reduced, the orbit of an ``S-"type cell
is shifted towards the orbits of ``A", with which the
rate of switch to ``A" is enhanced.  
The information of the cell-type distribution is represented by
the internal dynamics of ``S"-type cells, and it is essential to the regulation of 
differentiation rate \cite{FK98a}.

But why is the regulation oriented to keep the stability,
instead of the other direction?
In the example of Fig. 2,
the differentiation from $S$ to $A$ is enhanced when the type A cell is
removed.  Assume that the regulation worked in the other way 
( i.e., the decrease of
the rate $S\rightarrow A$ by the removal of A). 
Then, at the initial stage when type-A cells are produced, their number would decrease to zero,
since there are fewer type-A cells at the moment.  Hence,
the type-A cell would not appear from the beginning.
In other words, only the cell types that have a regulation mechanism
to stabilize their coexistence with other types can appear in our developmental process.

It should be stressed that our dynamical differentiation
process is always accompanied by this kind of regulation process, without any
sophisticated programs implemented in advance.
This autonomous robustness provides a novel viewpoint to 
the stability of the cell society in multicellular organisms.

\section{Irreversibility}

Since each cell state is realized as a balance between internal dynamics
and interaction, one can discuss which part is more relevant to determine
the stability of each state.
In one limiting case, the state is an attractor as internal dynamics,
which is sufficiently stable and not destabilized by cell-cell interaction.
In this case, the cell state is called `determined' following the terminology in cell biology.
In the other limiting case, the state is totally governed by the
interaction, and by changing the states of other cells, the cell state in concern
is destabilized.  In this case, each cell state is highly dependent on the
environment or other cells.

Each cell type in our simulation generally lies between these two limiting cases.
To see such intra-inter nature of the determination explicitly,
one effective method is a transplantation experiment.
Numerically, such experiment is carried out
by choosing determined cells (obtained from the normal differentiation process)
and putting them into a different set of surrounding cells,
to make a cell society that does not appear through the normal course
of development.

When a differentiated and recursive cell is transplanted to another cell society,
the offspring of the cell keep the same type,
unless the cell-type distribution of the society is strongly biased
When a cell is transplanted into a biased society, differentiation from
a `determined' cell occurs.  For example, a homogeneous society consisting only of one 
determined cell type is unstable, and some cells start to switch
to a different type.  Hence,
the cell memory is preserved mainly in each individual cell,
but suitable inter-cellular interactions are also necessary to keep it.

Following the above argument, cell types can be classified into groups
according to the degree of determination (Fig.3):

\begin{itemize}

\item
(U):'Undifferentiated cell class'; Cells change their chemical compositions
by generations.

\item
 (S):'Stem cell class'; Cells either reproduce the same type or switch to
different types with some rate.

\item
(D) 'Determined cell class';   
Chemical compositions are preserved by the division\footnotemark.

\end{itemize}

The degree of determination increases as (U) $\rightarrow$ (S) $\rightarrow$ (D).
In the class (U), the cell is in a transient state as a dynamical system,
and changes its character (e.g., average chemical composition) in time 
(and  accordingly by division).  The class (S) is
a stable state, and is regarded as a kind of ``attractor" {\em once the interaction 
term is specified}.
Still, the type is weak against the change of interaction, and can switch to a
different state, as mentioned.  
In the case of (ideally) determined type, it is an attractor by the internal 
dynamics itself.  In fact, when the interaction term is changed to
a large extent, it loses the stability, and switches to (U) or goes to a dead state.

In a real organism, there is a hierarchy in determination, and a stem cell is often 
over a progenitor over only a limited range of cell types.
In other words, the degree of determination is also hierarchical.  In our model,
we have also found such hierarchical structure.  So far, we have found only 
up to the second layer of hierarchy in our model with the number of chemicals $k=20$,
as shown in Fig.2.

In a real organism and in our model, there is a clear 
temporal flow, as (U) $\rightarrow$ (S) $\rightarrow$ (D),
as the development progresses.  The degree of determination increases in the normal course of
development.  Then how can one quantify such irreversibility?
Can we define some quantity similar to thermodynamic entropy?
So far no decisive answer is  available.  In our model simulation,
with the process of (U) $\rightarrow$ (S) $\rightarrow$ (D) in time

\begin{itemize}

\item
(I) Stability of intra-cellular dynamics increases

\item
(II) Diversity of chemicals decreases

\item
(III) Dynamics become less chaotic.

\end{itemize}

The degree of (I) could be determined by a minimum change 
in the interaction to switch a cell state, by properly extending the
`attractor strength' introduced in \cite{KK-Milnor}. A cell of (U) spontaneously
changes the state without the change of the interaction term, while the state of
(S) can be switched by tiny change in the interaction term.  The degree of determination in (D) 
is roughly measured as the minimum perturbation strength required for a
switch to a different state.

The diversity of chemicals (II) can be measured, for example, by $-\sum_{j=1}^k p(j) log p(j)$, with
$p(j)=<\frac{x(j)}{\sum_{m=1}^k x(m)}>$, with $<..>$ as temporal average.  For the example of Fig.2,
this index of diversity is 2.8 for type S, 1.4 for A, 2.4 for B, 1.0 for
A3 and A4, .9 for A5.  Successive decrease as S $\rightarrow$ A, and $\rightarrow$ A1,A2, and A3
is numerically expressed. 

The tendency (III) is numerically confirmed by the subspace Kolmorogorv-Sinai (KS) entropy
of the internal dynamics for each cell.  Here, this subspace KS entropy is measured as
a sum of positive Lyapunov exponents, in the tangent space restricted only to the
intracellular dynamics for a given cell.   In the example of Fig.2, the KS entropy is
positive ($\sim$ .001) for S, while for others it is almost zero.

\section{Discussion}

In the present paper we have provided
a new look at developmental cell biology, based on the
intra-inter dynamics, and production.
Differentiated cell types are given as stable states
realized as a balance between internal dynamics and interaction.
Hierarchical differentiation starting from a stem cell is a general 
consequence of our scenario.  Stochastic differentiation
with spontaneous regulation of differentiation probabilities
is found to be supported by intra-cellular chaotic dynamics.
In the present paper we claim that
cell differentiation and the formation of cell societies
are a natural  consequence of 
reproducing interacting units with internal dynamics.
It is just a universal phenomenon in a system having
such intra-inter dynamics.

Stability and irreversibility in development are stressed as a
universal feature in a developmental system with intra-inter dynamics.
The irreversibility is proposed to be expressed as the increase of
stability of each cellular state, and decrease of diversity in chemicals
and KS entropy.

We have also applied the idea to evolution, which shows that the
interaction-induced phenotypic differentiation is later fixed to genes.
Distinct species with distinct genotypes and phenotypes are formed
from a population of organisms with identical genotypes.
This leads to a scenario for robust sympatric speciation\cite{KY99}.

{\sl acknowledgments}

We are grateful to Tetsuya Yomo for fruitful discussion through collaborated
studies.  We would also like to thank stimulating discussions with Shin'ichi 
Sasa and Takashi Ikegami.
The work is supported by Grants-in-Aids for Scientific
Research from the Ministry of Education, Science, and Culture of Japan.

\pagebreak
\begin{figure}
\noindent
\hspace{-.3in}
\epsfig{file=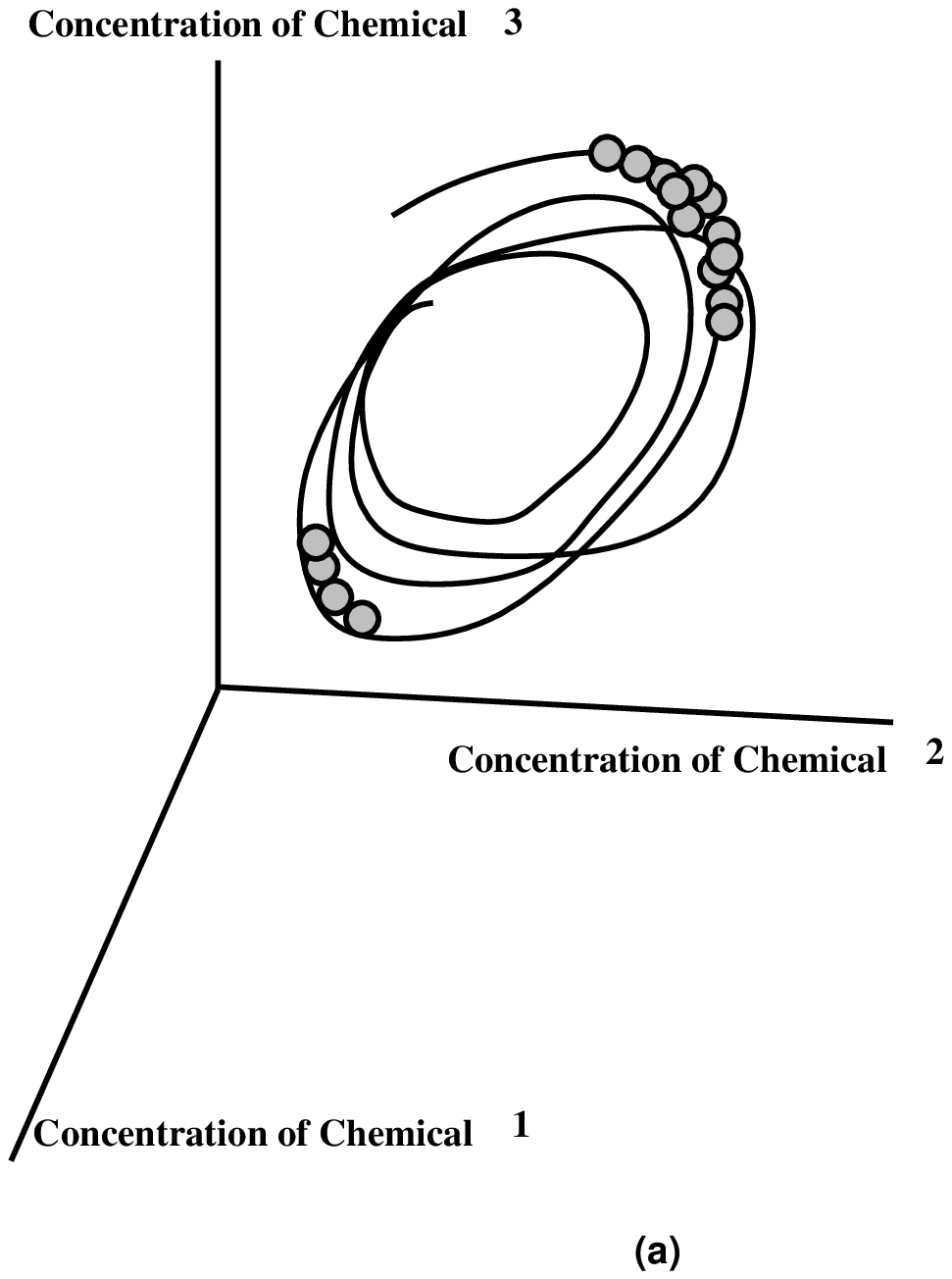,width=.45\textwidth}
\epsfig{file=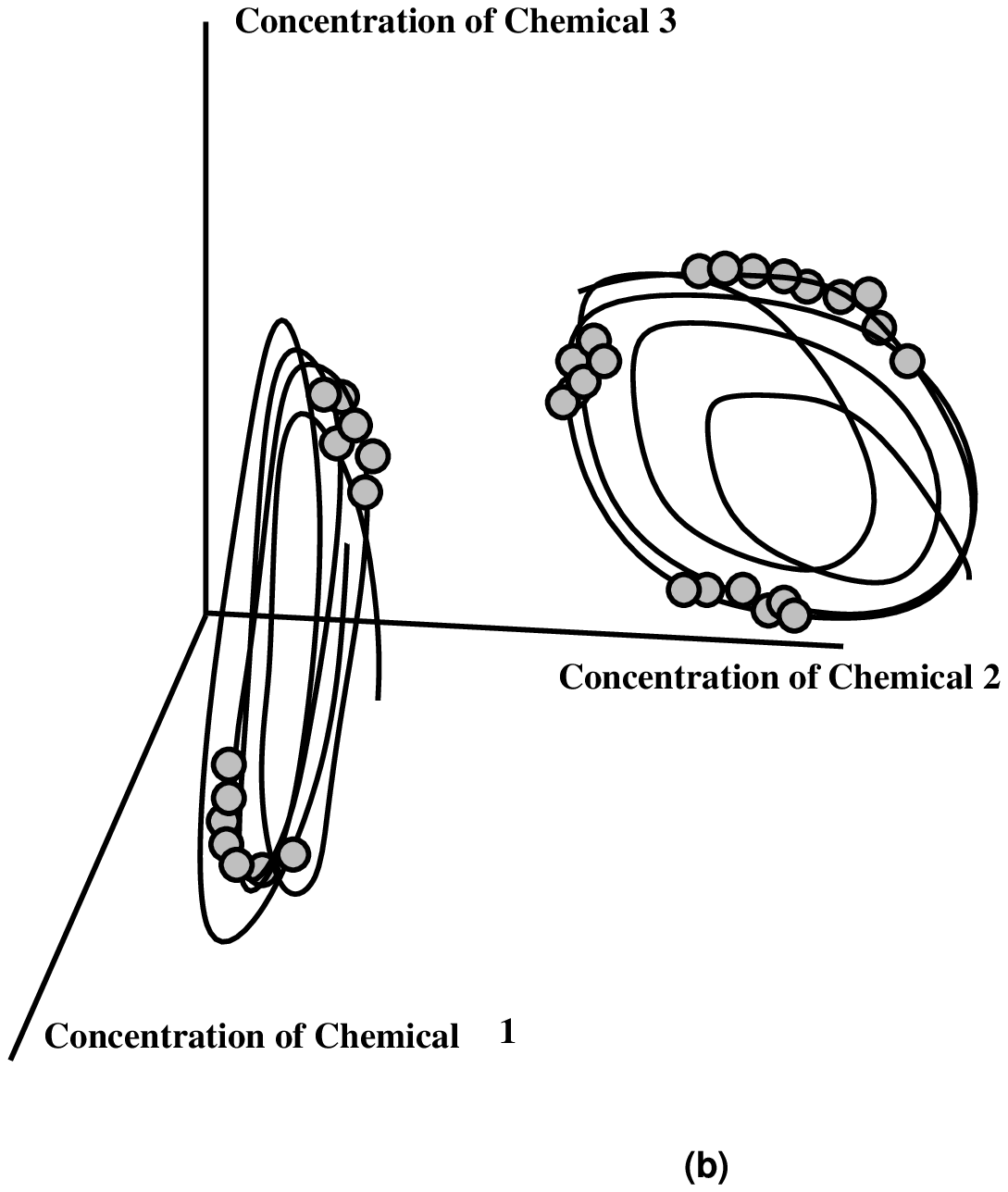,width=.45\textwidth}
\epsfig{file=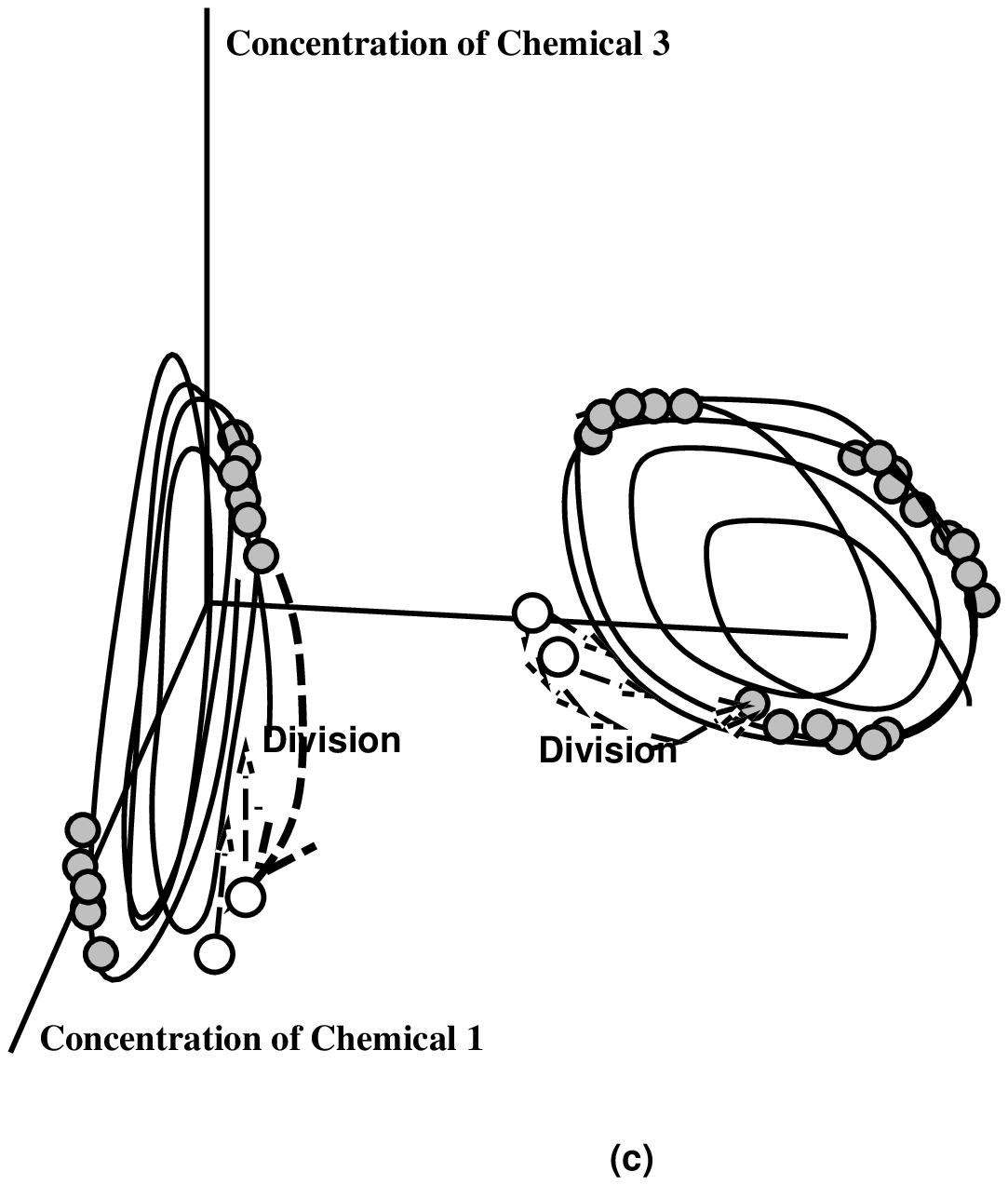,width=.45\textwidth}
\caption{Schematic representation of the stages of our differentiation scenario.
These results are supported by numerical simulations of several models.
See Refs. [5] and [8] for direct numerical results. }
\end{figure}

\pagebreak
\begin{figure}
\noindent
\hspace{-1.85in}
\epsfig{file=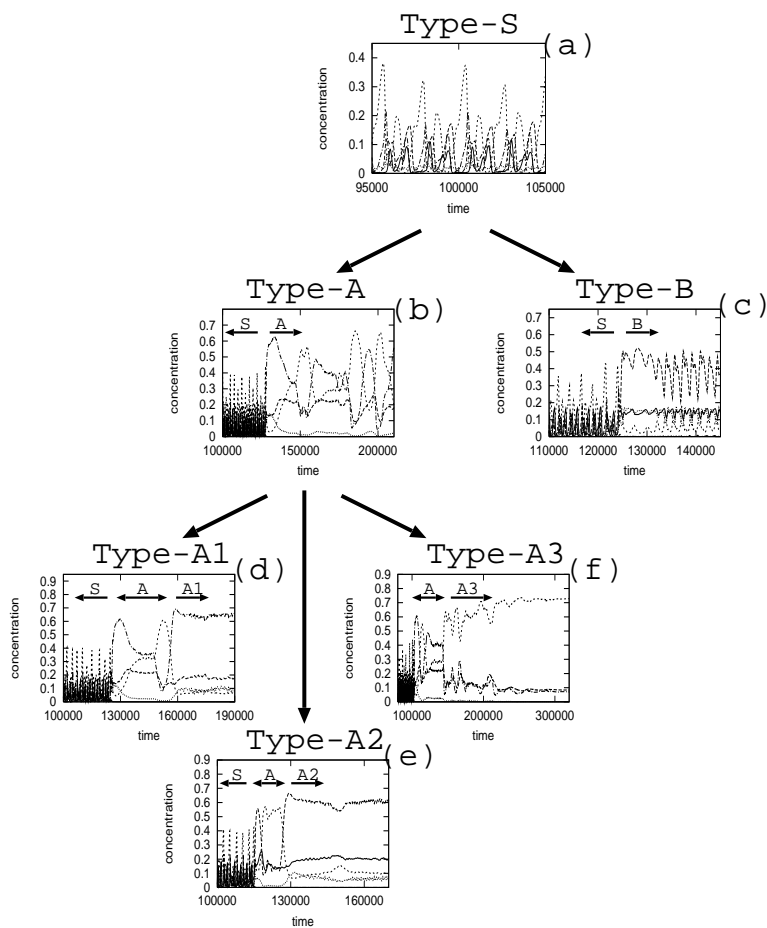,width=\textwidth}
\caption{ An example of  hierarchical differentiation observed in our model:
Fig.(a) represents the time series of chemical concentration of the type-S(stem cell).
Figs.(b)-(f) represent the course of differentiation to the corresponding cell type.
In this example, we use a network with 20 chemicals and the time series of 5 chemicals are plotted,
with different line styles.  Starting from a single cell, the type S appears initially. 
As the number increases the differentiations represented by arrows progress successively.
The type-S has a potential to differentiate to type-A,B and type-A can further differentiate to type-A1,A2,A3.
See also Ref.[8].}
\end{figure}

\pagebreak
\begin{figure}
\noindent
\hspace{-.3in}
\epsfig{file=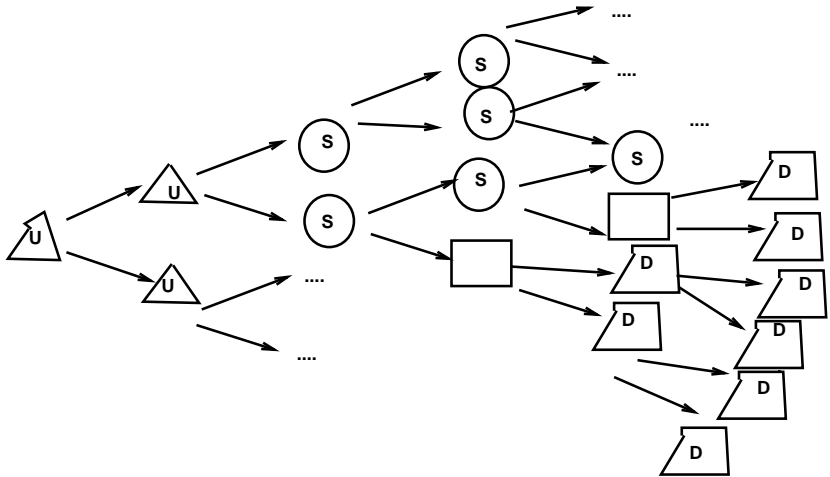,width=.8\textwidth}
\caption{ Schematic representation of differentiation of cell types.}
\end{figure}

\end {document}